\documentstyle[aps,prl]{revtex}

\begin{document}
\draft
\title{Quantum Langevin theory of excess noise}
\author{P.\ J.\ Bardroff and S.\ Stenholm}
\address{Department of Physics, Royal Institute of
  Technology (KTH),
  Lindstedtsv\"agen 24, S-10044 Stockholm, Sweden
}
\date{\today}
\maketitle
\begin{abstract}
In an earlier work [P.\ J.\ Bardroff and S.\ Stenholm, submitted to
Phys.\ Rev.\ Lett.], we
have derived a fully quantum mechanical description of excess noise in
strongly damped lasers.
This theory is used here to derive the corresponding quantum Langevin
equations.
Taking the semi-classical limit of these we are able to regain the
starting point of Siegman's treatment of excess noise [Phys.\ Rev.\ A
{\bf 39}, 1253 (1989)].
Our results essentially constitute a
quantum derivation of his theory and allow some generalizations.
\end{abstract}

\section{Introduction}
\label{sec:intro}

According to the fluctuation-dissipation theorem, the attenuation
or amplification of a signal always adds noise.
In optical amplifiers, this fact is usually phrased as ``one noise
photon'' added to the signal from the spontaneous emission processes
in the reservoir.
This assumption about the noise gives rise to the phase diffusion
responsible for the Schawlow-Townes
linewidth \cite{schawlow,scully,lax} of lasers.
However, this is not true generally.
In particular cases, the noise can exceed the intensity of one
photon by the so-called excess-noise factor or Petermann $K$-factor
\cite{peter}.

Experimentally this phenomenon was first confirmed in a laser cavity with
large output coupling leading to an enhancement of a few times
\cite{output}.
Later, even a factor of a few hundreds was achieved for solid state lasers
\cite{semicond} and gas 
lasers \cite{gas}.
Also experiments with a coupling of the
polarizations \cite{polar} 
and an inserted small aperture \cite{aperture} have demonstrated large
excess noise. 
A recent experiment has shown that excess noise can be colored
due to saturation effects \cite{color}.

After the prediction of excess noise by Petermann for the case of
gain-guided semiconductor lasers \cite{peter}, the first more general
theory of excess noise was given by Siegman using a
semi-classical description \cite{semi}.
Until recently, only a few simple systems have been discussed from a
quantum mechanical point of view \cite{squant}.

In a previous paper \cite{bardroff} we introduced a master equation
describing a multi-mode field interacting with a reservoir describing the
general linear amplifier or attenuator in a strictly quantum
mechanical formulation.
We find that under certain conditions, the reservoirs
create couplings between the undamped modes of the system.
Such dissipative couplings lead to a non-Hermitian eigenvalue problem,
which introduces non-orthogonal quasi modes in a natural manner.
The amplitudes of these modes are then found to display the expected
excess noise, which we here ascribe to the reservoir-induced mode-mode
coupling.

In this paper we derive the quantum Langevin
formalism following from our theory in Ref.~\cite{bardroff}.
Whereas the dynamic variable of the master equation is the quantum
state, the Langevin equations are for the field operators.
This allows us a direct comparison of our approach with the well-known
semi-classical treatment introduced by Siegman \cite{semi}.
As this has provided the physical understanding and the mathematical
expressions for the excess noise, we are pleased that we can
essentially derive his starting equations from our fully quantum
mechanical treatment.
We are also able to generalize the semi-classical analysis of excess
noise to cases beyond the paraxial approximation.

\section{Master equation}
\label{sec:master}

In this section we briefly review the results of the quantum
derivation of the excess-noise factor based upon the master equation.
We use orthonormal real mode functions $u_n(x)$ of the electromagnetic
field with frequency $\omega_n$ which fulfill the boundary conditions
for the given configuration in the whole ``universe'' and satisfy the
orthonormality relation
\begin{equation}
  \frac{1}{V}\int d^3x\,u_n(x)u_m(x)=\delta_{nm},
  \label{eq:orthu}
\end{equation}
where $V$ is the volume of the whole space.
Note that the mode function $u_n(x)$ is a vector including the
polarization orientation and that we choose them to be real for
convenience.
The electric field operator then reads
\begin{equation}
  \label{eq:efield}
  \hat E(x)=\sum_n \varepsilon_nu_n(x)\left(\hat a_n +\hat
    a_n^\dagger\right),
\end{equation}
where $\hat a_n$ and $\hat a_n^\dagger$ are the usual creation and
annihilation operators of the field excitations and the so-called vacuum
field amplitude is
\begin{equation}
  \varepsilon_n=\sqrt{\frac{\hbar \omega_n}{2\epsilon_0 V}}.
\label{eq:vacfield}
\end{equation}

We start from the multi-mode master equation \cite{bardroff}
\begin{eqnarray}
  \frac{d}{dt}\hat\rho(t)&=&\frac{1}{2}\sum_{n,m}L_{m,n}\left\{
    2\hat a_n^\dagger\hat\rho(t)\hat a_m - \hat a_m\hat
    a_n^\dagger\hat\rho(t) - \hat\rho(t)\hat a_m\hat a_n^\dagger
  \right\}\nonumber\\
  &&+\frac{1}{2}\sum_{n,m}\Gamma_{m,n}\left\{
    2\hat a_n\hat\rho(t)\hat a_m^\dagger - \hat a_m^\dagger\hat
    a_n\hat\rho(t) - \hat\rho(t)\hat a_m^\dagger\hat a_n
  \right\}
  -i\sum_n\omega_n[\hat a_n^\dagger\hat a_n,\hat\rho(t)].
  \label{eq:masterda}
\end{eqnarray}
with the two symmetric matrices $\Gamma_{m,n}$ and $L_{m,n}$ given by
\begin{mathletters}
\begin{equation}
  \label{eq:Gamma}
  \Gamma_{m,n}=\frac{\tau^2}{\hbar^2} \varepsilon_n\varepsilon_m
  \frac{1}{V}\int d^3x\, r_\Gamma(x)[u_n(x)d][u_m(x)d]
\end{equation}
and
\begin{equation}
  L_{m,n}=\frac{\tau^2}{\hbar^2} \varepsilon_n\varepsilon_m
  \frac{1}{V}\int d^3x\, r_L(x)[u_n(x)d][u_m(x)d].
  \label{eq:L}
\end{equation}
\label{eq:GaLa}
\end{mathletters}
The former describes losses and the latter amplification due to the
interaction with the reservoirs.

The two reservoirs for amplification and attenuation are assumed to
consist of two-level atoms injected in the upper or lower state,
respectively \cite{scully}. 
They are completely characterized by the
position dependent injection rates $r_L(x)$ and $r_\Gamma(x)$, the
interaction time $\tau$ of the individual atoms with the field and the
orientation of the atomic dipole moment $d$.
In principle, the dipole orientation could be different for damping and
attenuation and it may depend on position.
This treatment of the damping can describe spatially localized
absorption due to an inserted aperture or due to a detector placed
outside the cavity.
Assuming a perfect absorber (or detector) surrounding our
cavity, the reservoir can also model the damping due to output coupling.
Taking the limit of a infinitely large ``universe''
($V\rightarrow\infty$), and hence using a
continuum of modes, would be
another way of including losses due to output coupling in our model
as shown in Ref.~\cite{lang}.

Because of the interaction through the reservoir, the time evolution
of the mean values 
\begin{equation}
  \label{eq:meana}
  \frac{d}{dt}\langle\hat a_n\rangle=\frac{1}{2}\sum_m\left(
    L_{m,n}-\Gamma_{m,n}\right)\langle\hat a_m\rangle
  -i\omega_n\langle\hat a_n\rangle
\end{equation}
exhibits coupling between different modes. 
The definition of the quasi modes operator $\hat A$ follows from
imposing the condition 
\begin{equation}
  \label{eq:qmodeA}
  \frac{d}{dt}\langle \hat A\rangle
  =\Big\{\frac{1}{2}(\lambda-\gamma)-i\Omega\Big\}\langle \hat A \rangle,
\end{equation}
where $\Omega$ is the frequency, $\lambda$ is the
amplification rate and $\gamma$ is the attenuation rate.
For later convenience, we split the net-amplification rate
$(\lambda-\gamma)$ into the two separate contributions $\lambda$ and
$\gamma$.  Note that $\Omega$, $\lambda$ and $\gamma$ are real.
We write this mode operator in terms of the free field mode
operators as
\begin{equation}
  \label{eq:decomp}
  {\cal E}\hat A=\sum_n\varepsilon_n c_n \hat a_n
\end{equation}
with
the expansion coefficients $c_n$.
This transformation includes the vacuum-field amplitudes
$\varepsilon_n$ and  we define
${\cal E}=\sqrt{\frac{\hbar\Omega}{2\epsilon_0 V}}$,
because then the classical field amplitudes
$\varepsilon_n\langle \hat a_n\rangle$ obey the same transformation as
the operators.
Inserting Eq.~(\ref{eq:qmodeA}) into (\ref{eq:decomp}) we get an
eigenvalue equation
\begin{equation}
  \label{eq:eigenvalue}
  \sum_n\left\{ \frac{1}{2}(L_{m,n}-\Gamma_{m,n})
    -i\delta_{n,m}\omega_n \right\}
  \frac{\varepsilon_n}{\varepsilon_m} c_n=
  \left\{\frac{1}{2}(\lambda-\gamma)-i\Omega\right\} c_m
\end{equation}
for the non-Hermitian matrix $\{\frac{1}{2}(L_{m,n}-\Gamma_{m,n})
-i\delta_{n,m}\omega_n\} \frac{\varepsilon_n}{\varepsilon_m}$.
Here $c_n^{(\nu)}$ is the right eigenvector;
the corresponding left eigenvector is $\varepsilon^2_n c_n^{(\nu)}$
\cite{fn:ev}.
The superscript $\nu$ distinguishes the different eigenvectors.
The detailed properties of the quasi modes are summarized in the
Appendix.

We can now calculate the noise of the quadrature operator
\begin{equation}
  \label{eq:X}
  \hat X_\nu(x)=
  {\cal E}_\nu
  \left[U_\nu(x)\hat A_\nu+U^*_\nu(x)\hat A_\nu^\dagger\right]
\end{equation}
with the definition of the quasi-mode function $U_\nu(x)$ given by
Eq.~(\ref{eq:rmode}) in the Appendix. 
Taking the noise averaged over position and comparing to the usual single
mode master equation with the same frequency $\Omega_\nu$, damping rate
$\gamma_\nu$ and amplification rate $\lambda_\nu$ we find an
enhancement by the factor
\begin{equation}
  \label{eq:Kqm}
  K_\nu=\left|\frac{\sum\varepsilon_n^2|c^{(\nu)}_n|^2}{
      \sum_m \varepsilon_m^2
      {c^{(\nu)}_m}^2}\right|^2
\end{equation}
for the noise added by the reservoir; cf.\ Ref.~\cite{bardroff}.
The excess noise is large when the matrices $L_{m,n}$ and
$\Gamma_{m,n}$, defined in Eqs.~(\ref{eq:GaLa}), have large off-diagonal
terms and when they are not identical.
The former follows when the injection rates $r_L(x)$ and $r_\Gamma(x)$
are not spatially constant whereas the latter when damping and
amplification are spatially separated.

\section{Quantum Langevin equation}
\label{sec:qlangevin}

Following the usual treatment \cite{lax}, we replace the time
evolution described by the master equation (\ref{eq:masterda}) by an
equivalent quantum Langevin equation.
This contains non-commuting noise forces which are designed such as to
give the same moments as those derived from the master equation.
The quantum Langevin equation is written
\begin{equation}
  \label{eq:qla}
  \frac{d}{dt}\hat a(t)_n=\frac{1}{2}\sum_m\left(
    L_{m,n}-\Gamma_{m,n}\right)\hat a_m(t)
  -i\omega_n\hat a_n(t)+\hat f_n(t),
\end{equation}
where the Langevin noise sources $\hat f_n(t)$ obey the correlation
relations
\begin{mathletters}
  \begin{eqnarray}
    \label{eq:corrdef}
    \langle \hat f_m(t) \hat f_n^\dagger(t')\rangle&=&2\langle\hat D_{\hat
      a_m \hat a_n^\dagger} \rangle\delta(t-t'),\\
    \langle \hat f_n^\dagger(t)\hat f_m(t')\rangle&=&2\langle\hat D_{
    \hat a_n^\dagger\hat a_m} \rangle\delta(t-t')
  \end{eqnarray}
\end{mathletters}
and
\begin{equation}
  \langle \hat f_m(t)\rangle=
  \langle \hat f_m(t) \hat f_n(t')\rangle=\langle \hat
  f_m^\dagger(t) \hat f_n^\dagger(t')\rangle=0.
  \label{eq:corrzero}
\end{equation}
It then follows from Eq.~(\ref{eq:qla}) that the expectation values
obey Eq.~(\ref{eq:meana}).
The diffusion coefficients $\langle\hat D_{\hat a_m \hat
  a_n^\dagger}\rangle$ and 
$\langle\hat D_{\hat a_n^\dagger\hat a_m} \rangle$ have to be
determined to give the correct noise correlations $\langle\hat a_n^\dagger
  \hat a_m\rangle$ and $\langle\hat a_m\hat a_n^\dagger\rangle$.
We compare the time evolution of the noise correlations derived
from the master equation, given by Eqs.~(\ref{eq:acorr}),
to the one derived from the Langevin equation (\ref{eq:qla}) to find
the relations 
\begin{mathletters}
  \label{eq:corrcorr}
  \begin{equation}
    \langle\hat f_n^\dagger(t)\hat a_m(t)
    +\hat a_n^\dagger(t)\hat f_m(t)\rangle=\Gamma_{m,n}
  \end{equation}
and
\begin{equation}
    \langle\hat f_n(t)\hat a_m^\dagger(t)
    +\hat a_n(t)\hat f_m^\dagger(t)\rangle=L_{m,n}.
  \end{equation}
\end{mathletters}
With the help of the Einstein relations 
\begin{mathletters}
  \label{eq:einstein}
  \begin{eqnarray}
     2\langle\hat D_{\hat a_m \hat a_n^\dagger}\rangle&=&\frac{d}{dt}
     \langle\hat a_m(t)\hat a_n^\dagger(t)\rangle-
     \langle\hat a_m(t)
(\frac{d}{dt}\hat a_n^\dagger(t)-\hat f_n^\dagger(t))\rangle-
     \langle(\frac{d}{dt}\hat a_m(t)-\hat f_m(t))
\hat a_n^\dagger(t)\rangle,
\\
     2\langle\hat D_{\hat a_n^\dagger\hat a_m}\rangle&=&\frac{d}{dt}
     \langle\hat a_n^\dagger(t)\hat a_m(t)\rangle-
     \langle(\frac{d}{dt}\hat a_n^\dagger(t)-\hat f_n^\dagger(t))
\hat a_m(t)\rangle-
     \langle\hat a_n^\dagger(t)
(\frac{d}{dt}\hat a_m(t)-\hat f_m(t))\rangle,
  \end{eqnarray}
\end{mathletters}
and Eqs.~(\ref{eq:corrcorr}), we
determine the diffusion coefficients to be
\begin{mathletters}
  \label{eq:noisecorr}
  \begin{equation}
    2\langle\hat D_{\hat a_m \hat a_n^\dagger}\rangle=
    \langle\hat f_n^\dagger(t)\hat a_m(t)
    +\hat a_n^\dagger(t)\hat f_m(t)\rangle=\Gamma_{m,n}
  \end{equation}
and
\begin{equation}
  2\langle\hat D_{\hat a_n^\dagger\hat a_m} \rangle=
    \langle\hat f_n(t)\hat a_m^\dagger(t)
    +\hat a_n(t)\hat f_m^\dagger(t)\rangle=L_{m,n}.
  \end{equation}
\end{mathletters}
These relations clearly show the mode correlations due to the reservoir.

In the following, we derive a wave equation for the propagation of the
electric field 
operator including amplification, damping and the corresponding noise
source.
Starting from Eqs.~(\ref{eq:qla}) and (\ref{eq:efield}) we can find
the exact equation
\begin{eqnarray}
  \label{eq:exact}
  \left\{\frac{\partial^2}{\partial t^2}
    -c^2\nabla^2
  \right\} \hat E(x,t)-\sum_n\varepsilon_n u_n(x)\Big\{
  \sum_k(L_{k,n}-\Gamma_{k,n})\frac{d}{dt}\hat a_k+h.c.\Big\}
  =\nonumber\\
  \sum_n\varepsilon_n u_n(x)\Big\{
  -\frac{1}{4}\sum_{k,l}(L_{k,n}-
  \Gamma_{k,n})(L_{l,k}-\Gamma_{l,k})\hat a_l\nonumber\\
  +\frac{i}{2}\sum_k
  (L_{k,n}-\Gamma_{k,n})(\omega_k-\omega_n)\hat a_k\nonumber\\
  -\frac{1}{2}\sum_k(L_{k,n}-\Gamma_{k,n})\hat f_k-i\omega_n\hat f_n
  +\frac{d}{dt}\hat f_n
  \Big\}
  +h.c.
\end{eqnarray}
with the mode functions $u_n(x)$ fulfilling the Helmholtz equation
\begin{equation}
  \label{eq:hh}
  (c^2\nabla^2 +\omega_n^2)u_n(x)=0
\end{equation}
together with the appropriate boundary conditions.
At this point we introduce a number of approximations based on the
assumption that the average oscillation frequency $\bar\omega$ of the
electric field is much higher than the decay or amplification rates,
e.g.\ $|L_{n,m}|$ or $|\Gamma_{n,m}|$.
This is well justified in the optical regime where the former
is at least six orders of magnitude larger than the latter.
The spectral width $\Delta\omega$ of the relevant frequencies $\omega_n$ is
assumed to be of the order of the decay or amplification rate.
To be more specific, we will neglect terms of the order 
${\cal O}(\lambda_\nu/\bar\omega)^2$, 
${\cal O}(\lambda_\nu \Delta\omega/\bar\omega^2)$ and
${\cal O}(\Delta\omega/\bar\omega)^2$ or smaller, and
we assume
${\cal O}(\lambda_\nu)\approx {\cal O}(\gamma_\mu)$.

On the LHS of Eq.~(\ref{eq:exact}) the two terms 
$\frac{\partial^2}{\partial t^2}\hat E(x,t)$ and 
$c^2\nabla^2 \hat E(x,t)$ are of the order ${\cal O}(\bar\omega)^2$.
Since the remaining term on the LHS of Eq.~(\ref{eq:exact}) is
proportional to the damping and amplification rate, we can approximate
the frequencies by the mean frequency $\bar \omega$.
Hence inserting the definitions of $L_{n,m}$ and $\Gamma_{n,m}$,
Eqs.~(\ref{eq:GaLa}), and of $\varepsilon_n$, Eq.~(\ref{eq:vacfield}),
the remaining term on the LHS of Eq.~(\ref{eq:exact}) yields
\begin{eqnarray}
  \label{eq:approxLG}
  &&\sum_n\varepsilon_n u_n(x)
  \sum_k(L_{k,n}-\Gamma_{k,n})\frac{d}{dt}\hat a_k+h.c.=\nonumber\\
  &&\sum_n\varepsilon_n u_n(x)\sum_k\frac{\tau^2}{\hbar^2}
  \varepsilon_n\varepsilon_k
  \frac{1}{V}\int d^3x'(r_L(x')-r_\Gamma(x'))[u_n(x')d][u_k(x')d]
  \frac{d}{dt}\hat a_k+h.c.=\nonumber\\
  &&\frac{\tau^2}{\hbar^2}\sum_n\varepsilon_n^2 u_n(x)
  \frac{1}{V}\int d^3x'(r_L(x')-r_\Gamma(x'))[u_n(x')d]d^T
  \frac{\partial}{\partial t}\hat E(x',t)
  \approx\nonumber\\
  &&\frac{\tau^2 \bar\omega}{2\epsilon_0\hbar}\int
  d^3x'(r_L(x')-r_\Gamma(x'))\delta_T(x-x')d\otimes d^T
  \frac{\partial}{\partial t}
  \hat E(x',t)=
  (R_L(x)-R_\Gamma(x))
    \frac{\partial}{\partial t}\hat E(x,t).
\end{eqnarray}
Here the matrices $L_{n,m}$ and $\Gamma_{n,m}$ occur in their position
representations
\begin{mathletters}
  \label{eq:posrep}
  \begin{equation}
   \frac{1}{V}
    \sum_{n,m}u_n(x)\otimes u_m^T(x')L_{n,m}\approx
    \frac{\tau^2 \bar\omega}{2\epsilon_0\hbar}
    r_L(x)\delta_T(x-x')d\otimes d^T\equiv
     R_L(x)\delta_T(x-x')
  \end{equation}
and
  \begin{equation}
    \frac{1}{V}
    \sum_{n,m}u_n(x)\otimes u_m^T(x')\Gamma_{n,m}\approx
    \frac{\tau^2 \bar\omega}{2\epsilon_0\hbar}
    r_\Gamma(x)\delta_T(x-x')d\otimes d^T\equiv
    R_\Gamma(x)\delta_T(x-x').
  \end{equation}
\end{mathletters}
Note that $R_L(x)$, $R_\Gamma(x)$, $d\otimes d^T$ and the transverse
$\delta$-function
\begin{equation}
  \delta_T(x-x')=\frac{1}{V}\sum_n u_n(x)\otimes u_n^T(x')
  \label{eq:delta}
\end{equation}
are tensors.
We can neglect the terms on the RHS of Eq.~(\ref{eq:exact})
containing the field operator
since they are of the order ${\cal O}(\lambda_\nu)^2$ or
${\cal O}(\lambda_\nu \Delta\omega)$, respectively.
For the noise we only take terms of lowest order.
Therefore we may neglect the first of the noise terms in
Eq.~(\ref{eq:exact}) and we approximate $\frac{d}{dt}\hat
f_n\approx-i\bar\omega\hat f_n$.
Introducing the position representation
\begin{equation}
  \label{eq:fpos}
  \hat f(x,t)=\sum_n \varepsilon_n u_n(x) \hat f_n(t)
\end{equation}
of the noise source we find
\begin{equation}
  \label{eq:tele}
  \left\{\frac{\partial^2}{\partial t^2}
    -c^2\nabla^2-(R_L(x)-R_\Gamma(x))
    \frac{\partial}{\partial t}
  \right\} \hat E(x,t)=-2i\bar\omega\hat f(x,t) +h.c.
\end{equation}
The correlations of the noise operators are
\begin{mathletters}
  \label{eq:corrpos}
  \begin{equation}
    \langle \hat f(x,t) \hat f^\dagger(x',t')\rangle=
    \frac{\hbar\bar\omega}{2\epsilon_0}R_\Gamma(x)
    \delta_T(x-x')\delta(t-t')
  \end{equation}
and
  \begin{equation}
    \langle \hat f^\dagger(x,t) \hat f(x',t')\rangle=
    \frac{\hbar\bar\omega}{2\epsilon_0}R_L(x)
    \delta_T(x-x')\delta(t-t').
  \end{equation}
\end{mathletters}
Consequently the total noise on the RHS of Eq.~(\ref{eq:tele}) obeys
\begin{equation}
  \label{eq:corrpostot}
  \langle(-2i\bar\omega\hat f(x,t)+h.c.)^2\rangle=
  \frac{\hbar\bar\omega^3}{\epsilon_0}(R_L(x)+R_\Gamma(x))
  \delta_T(x-x')\delta(t-t').
\end{equation}
As expected, the effects of amplification and damping add for the noise 
whereas they subtract for the amplification.

\section{Semi-classical treatment}
\label{sec:semi}

Starting from Eq.~(\ref{eq:tele}) we can now perform the transition to the
semi-classical treatment replacing operators with $c$-numbers.
The solution of Eq.~(\ref{eq:tele}) can conveniently be written using
the positive frequency part $E^{(+)}$ of the electromagnetic field.
The real part is the electric field and the
imaginary part relates to the magnetic field.
With the help of the Green function
\begin{equation}
  \label{eq:green}
  G^{(+)}(x,x',t)=\sum_\nu U_\nu(x)\bar U_\nu(x')
  e^{\frac{1}{2}(\lambda_\nu-\gamma_\nu)t-i\Omega_\nu t}
\end{equation}
and the accumulated noise
\begin{equation}
  \label{eq:addnoise}
  N^{(+)}(x,t)=-2i\bar\omega \int\limits_0^t\! dt' \sum_\nu
  e^{\frac{1}{2}(\lambda_\nu-\gamma_\nu)(t-t')-i\Omega_\nu (t-t')}
  U_\nu(x)\frac{1}{V}\int d^3x'\bar U_\nu(x') f(x',t'),
\end{equation}
we find the field to be given by
\begin{equation}
  \label{eq:evol}
  E^{(+)}(x,t)=\frac{1}{V}\int d^3x' G^{(+)}(x,x',t) E^{(+)}(x',0)
  +N^{(+)}(x,t)
\end{equation}
starting from the initial field $E^{(+)}(x',t=0)$.
Within the approximations made, the quasi-mode functions $U_\nu(x)$ and
$\bar U_\nu(x)$ and their eigenvalues $\lambda_\nu$, $\gamma_\nu$ and
$\Omega_\nu$ are the same as defined using the master equation,
Eqs.~(\ref{eq:rmode}) and (\ref{eq:lmode}). 
When we now calculate the variance of the electric field $E(x,t)$
averaged over position and compare with
damping and amplification processes described by the usual single mode
master equation, we recover the same
$K$-factor, Eq.~(\ref{eq:Kqm}), as before.
We find for the noise term
\begin{eqnarray}
  \label{eq:Nsq}
  \frac{1}{V}\int d^3x \langle (N^{(+)}(x,t)+N^{(-)}(x,t))^2\rangle
  =\nonumber\\
  \frac{\hbar\bar\omega^3}{\epsilon_0}\sum_{\nu,\mu}\frac{
    \exp\{\frac{1}{2}(\lambda_\nu+\lambda_\mu-\gamma_\nu-\gamma_\mu)t-
    i(\Omega_\nu-\Omega_\mu)t\}-1}{\frac{1}{2}(\lambda_\nu+
    \lambda_\mu-\gamma_\nu-\gamma_\mu)-i(\Omega_\nu-\Omega_\mu)}
  \nonumber\\
  \times
  \frac{1}{V}\int d^3x\,U_\nu(x)U_\mu^*(x)\frac{1}{V}
  \int d^3x\,\bar U_\nu(x)
  (R_L(x)+R_\Gamma(x))\bar U_\mu^*(x).
\end{eqnarray}
Considering only one quasi mode, the noise simplifies to
\begin{eqnarray}
  \label{eq:Nsq2}
  \frac{1}{V}\int d^3x \langle (N_\nu^{(+)}(x,t)+N_\nu^{(-)}(x,t))^2\rangle
  \approx\nonumber\\
  \frac{\hbar\bar\omega^3}{\epsilon_0}\frac{
    \exp\{(\lambda_\nu-\gamma_\nu)t\}-1}{\lambda_\nu-\gamma_\nu}
  (\lambda_\nu+\gamma_\nu)K_\nu
\end{eqnarray}
with the enhancement factor in the commonly used form
\begin{equation}
  \label{eq:KK}
  K_\nu=\frac{\int d^3x\,U_\nu(x)U_\nu^*(x) \int d^3x\,\bar U_\nu(x) \bar
    U_\nu^*(x)}{\left|\int
      d^3x\,U_\nu(x)\bar U_\nu(x)\right|^2}.
\end{equation}
Note that our choice of the normalization for the quasi-mode functions
is given as in Eq.~(\ref{eq:ortho}). 
We have shown in Ref.~\cite{bardroff} that Eq.~(\ref{eq:KK}) agrees
with Eq.~(\ref{eq:Kqm}) up to the order
${\cal O}(\Delta\omega/\bar\omega)^2$.

Siegman \cite{semi} used an equation analogous to Eq.~(\ref{eq:tele}) 
as the starting point for his derivation of the excess-noise factor.
However, there are two interesting differences in the details of the
noise source correlations, Eq.~(\ref{eq:corrpostot}).

The first difference in Ref.~\cite{semi} is that the spatial
transverse $\delta$-function is replaced by a usual $\delta$-function
and that the temporal $\delta$-function is replaced by the Hertzian
bandwidth $\Delta\omega/(2\pi)$ of the reservoir.
The latter circumstance is explained by using the Fourier
representation of our noise correlation in Eq.~(\ref{eq:corrpostot});
this leads to the same equation with $\delta(t-t')$ replaced by
$\delta(\omega-\omega')/2\pi$.
We then integrate with respect to $\omega$ and $\omega'$ over the
frequency bandwidth $\Delta\omega$ to obtain
\begin{eqnarray}
  \frac{1}{2}\left[\,\,\,
  \int\!\!\!\!\!\!\!\!
  \int\limits_{\bar\omega-\Delta\omega/2}^{\bar\omega+\Delta\omega/2}\! 
  \frac{d\omega  d\omega'}{2\pi}
  \delta(\omega-\omega') e^{-i(\omega t-\omega' t')}+c.c.\right]
  &=&\frac{\sin(\Delta\omega(t-t')/2)}{2\pi(t-t')}e^{-i\bar\omega
    (t-t')}+c.c.
  \nonumber\\
  &\approx&\frac{1}{2}\left[
    \frac{\Delta\omega}{2\pi} e^{-i\bar\omega (t-t')}+c.c.\right]
  \label{eq:herzb}
\end{eqnarray}
for $|t-t'|<(\Delta\omega)^{-1}$.
This approximation is reasonable when the mean frequency $\bar\omega$
of the noise is much larger then the bandwidth $\Delta\omega$.

The second difference in Ref.~\cite{semi} is that
$(R_L(x)+R_\Gamma(x))$ is replaced by
$2\left(\frac{R_L}{R_L-R_\Gamma}\right)(R_L-R_\Gamma)=2R_L$
with spatially constant
$R_L$ and $R_\Gamma$.
This simplification is justified only when averaging over the whole
volume V and when amplification and damping are balanced.
For the derivation of the $K$-factor which involves an
average over position, this is a valid replacement.
However, one has to be aware of the subtlety that only non-constant
$R_L(x)$ and $R_\Gamma(x)$ with $R_L(x)\neq R_\Gamma(x)$ lead to
non-orthogonal quasi modes and hence can give $K>1$.

It is interesting to note that within the paraxial approximation, we
obtain an equation analogous to the position representation of
Eq.~(\ref{eq:qla})---the starting point of our semi-classical analysis.
Making the ansatz
\begin{equation}
  \label{eq:ansatz}
  E^{(+)}(x,t)=e^{i\bar\omega(z/c-t)}\tilde E^{(+)}(x)
\end{equation}
with $\tilde E^{(+)}(x)$ slowly varying with respect to the longitudinal
coordinate $z$, we get from Eq.~(\ref{eq:tele})
\begin{equation}
  \label{eq:paraxial}
  c \frac{\partial}{\partial z}\tilde E^{(+)}(x)=
  \left\{\frac{1}{2}(R_L(x)-R_\Gamma(x))
    +\frac{i c^2}{2\bar\omega}\nabla_T^2\right\}\tilde E^{(+)}(x)
  +\tilde f(x)
\end{equation}
where $f(x,t)\approx e^{i\bar\omega(z/c-t)}\tilde f(x)$.
The time derivative $d/dt$ of Eq.~(\ref{eq:qla}) is replaced
by the derivative with respect the longitudinal coordinate $c
\partial/\partial z$ which is equivalent in a frame moving
with the electromagnetic wave.
The frequency part of Eq.~(\ref{eq:qla}) is replaced by the
transverse Laplacian.
Frequently, Eq.~(\ref{eq:paraxial}) is solved with mode functions of the
transverse Laplace equation, depending only parametrically on the
longitudinal coordinate.
This distinction between longitudinal and transverse coordinates leads
to a factorization of the $K$-factor into a longitudinal and a
transverse part.

\section{Discussion}
\label{sec:discussion}

In an earlier paper \cite{bardroff}, we derived the master equation
for a set of modes coupled to amplifying and attenuating reservoirs.
This introduces couplings between the undamped modes of the total
``universe'' and leads directly to the introduction of quasi modes,
which are found to exhibit the excess noise
described originally by Petermann \cite{peter}.

Our treatment has been carried out only in the linear regime so far.
This describes an amplifier or an attenuator, where the treatment is
most straightforward and the results display the most transparent
physical insight.
However, the saturation in an operating laser will need to be
considered, and we are for the moment carrying out such calculations,
which show the influence of the excess noise in the strong field
situation.

The best physical picture of this noise was provided by Siegman
\cite{semi}, who also supplied the quasi-mode expression for the
excess-noise factor.
This has then been used successfully to describe the experimental
findings \cite{output,semicond,gas,polar,aperture}.
In \cite{bardroff} we showed that our quantum mechanical approach
naturally provides an expression which is essentially identical with
Siegman's results.

Siegman, however, utilized a semi-classical Langevin approach, where
the noise forces were added {\em ad hoc} to the classical equations
for the amplitudes; the noise forces were then supplied with properly
chosen correlation properties, which was shown to imply the presence
of excess noise.
Because this approach has been found to give both a physically
attractive and theoretically justified description of the situation,
we find it interesting to connect that treatment to our quantum
approach in some detail.

In this paper we derive the quantum Langevin equations following from
our general master equation.
Here we utilize techniques known from quantum noise theory, and obtain
results that can be directly compared with the treatment of Siegman's,
when the semi-classical limit is taken.
Except for some minor differences, our resulting equations are
identical with those used by Siegman.
We thus claim that we have justified his formulation of the problem
from a more fundamental quantum mechanical point of view.
The differences found are either based on natural approximations or
obvious qualifications of the results as e.g.\ the introduction of the
transverse delta function in the noise correlations.
In addition, we have been able to generalize the theory to
situations outside the paraxial approximation.

The results of our treatment, however, have bearings beyond the
problem of excess noise in highly lossy cavities.
The approach is quite general, and in
addition to the Markov approximation we only need the rotating wave
approximation for the interaction with the reservoirs.
The master equation is then derived from first principles, and the
nonorthogonal quasi modes emerge in a natural manner.
The theory is fully general and may well be
applicable to other high loss physical systems as well.
For the moment we know of no observation that would show the
equivalent of the laser excess noise, but novel situations may soon
turn up.
The lively research activity in quantum information processing, atom
optics and novel measurement situations may provide potential
applications of the present theory.

The physics of our approach resides in the coupling of the undamped
modes through the reservoirs.
In such a situation, the only essential assumption
in our derivation is the Markovian approximation.
In highly damped systems, this may not necessarily hold, and the
introduction of memory effects in our theory has not been considered
so far.
Some features are, however, expected to survive, but also unexpected
complications may appear.
These questions remain to be investigated.

\appendix

\section{Properties of the quasi modes}
\label{sec:qmodes}

In this Appendix we recall from Ref.~\cite{bardroff} those properties
of the quasi modes which are relevant for the derivation of the excess
noise.
The only properties of the left and right eigenvectors of
non-Hermitian matrices which we need for our analysis are their
mutual orthogonality and completeness \cite{footnote}:
The eigenvectors fulfill the orthogonality condition
\begin{equation}
  \label{eq:orthogonal}
  \sum_n \varepsilon_n^2
  c^{(\nu)}_nc^{(\mu)}_n=\delta_{\nu,\mu}\sum_n \varepsilon_n^2
  {c^{(\nu)}_n}^2
\end{equation}
and the completeness relation
\begin{equation}
  \label{eq:complete}
  \sum_\nu \left(\frac{
    \varepsilon_n^2 c^{(\nu)}_n c^{(\nu)}_m
    }{
    \sum_{n'} \varepsilon_{n'}^2 {c^{(\nu)}_{n'}}^2}\right)=
  \delta_{n,m}
\end{equation}
with $\sum_{n'} \varepsilon_{n'}^2 {c^{(\nu)}_{n'}}^2\neq0$.
Therefore, we can uniquely define the set of quasi-mode operators as
\begin{eqnarray}
  \label{eq:qA}
  \hat A_\nu=\frac{1}{{\cal E}_\nu}\sum_n c^{(\nu)}_n
  \varepsilon_n \hat a_n
\end{eqnarray}
with the vacuum field amplitude
\begin{equation}
  {\cal E}_\nu=\sqrt{\frac{\hbar\Omega_\nu}{2\epsilon_0 V}}.
  \label{eq:qvac}
\end{equation}
The inverse transformation is
\begin{equation}
  \label{eq:a}
  \hat a_n=\varepsilon_n\sum_\nu \frac{c^{(\nu)}_n}{\sum_m \varepsilon_m^2
  {c^{(\nu)}_m}^2}
  {\cal E}_\nu \hat A_\nu.
\end{equation}
Consequently the positive frequency part of the electric field
operator is given by
\begin{equation}
  \label{eq:E-}
  \hat E^{(+)}(x)=\sum_n \varepsilon_n u_n(x)\hat a_n=
  \sum_\nu {\cal E}_\nu U_\nu(x) \hat A_\nu.
\end{equation}
The quasi-mode eigenfunctions
\begin{equation}
  \label{eq:rmode}
  U_\nu(x)=\sum_n \frac{\varepsilon_n^2
  c^{(\nu)}_n}{\sum_m \varepsilon_m^2
  {c^{(\nu)}_m}^2} u_n(x)
\end{equation}
satisfy an orthogonality relation
\begin{equation}
  \label{eq:ortho}
  \frac{1}{V}\int d^3x\, U_\nu(x)\bar U_\mu(x)=\delta_{\nu,\mu}
\end{equation}
with their adjoint quasi-mode functions
\begin{equation}
  \label{eq:lmode}
  \bar U_\nu(x)=\sum_n c^{(\nu)}_n u_n(x).
\end{equation}
The properties
\begin{equation}
  \Omega_\nu=\frac{\sum_n \varepsilon_n^2\omega_n |c^{(\nu)}_n|^2}{\sum_n
  \varepsilon_n^2|c^{(\nu)}_n|^2}=
  \frac{2\epsilon_0 V}{\hbar}
  \frac{\sum_n \varepsilon_n^4 |c^{(\nu)}_n|^2}{\sum_n
  \varepsilon_n^2|c^{(\nu)}_n|^2},
  \label{eq:Omega}
\end{equation}
\begin{mathletters}
\begin{equation}
  \lambda_\nu=\frac{\sum_{n,m} L_{n,m}\varepsilon_n\varepsilon_m
    c^{(\nu)*}_n c^{(\nu)}_m}{\sum_n\varepsilon_n^2|c^{(\nu)}_n|^2}
  \label{eq:lambda}
\end{equation}
and
\begin{equation}
  \gamma_\nu=\frac{\sum_{n,m} \Gamma_{n,m}\varepsilon_n\varepsilon_m
    c^{(\nu)*}_nc^{(\nu)}_m}{\sum_n\varepsilon_n^2|c^{(\nu)}_n|^2}
  \label{eq:gamma}
\end{equation}
\end{mathletters}
can be obtained from the real and imaginary parts of
Eq.~(\ref{eq:eigenvalue}) after taking the scalar product with the
vector $\varepsilon_m^2c^{(\nu)*}_m$.

From the master equation (\ref{eq:masterda}) follows the time evolution
of the noise correlations
\begin{mathletters}
  \label{eq:acorr}
  \begin{eqnarray}
    \label{eq:acorra}
    \frac{d}{dt}\langle\hat a_n^\dagger \hat a_m\rangle=
    \frac{1}{2}\sum_k\left(
    L_{k,n}-\Gamma_{k,n}\right)\langle a_k^\dagger \hat a_m\rangle
  +\frac{1}{2}\sum_k\left(
    L_{m,k}-\Gamma_{m,k}\right)\langle a_n^\dagger \hat a_k\rangle
  \nonumber\\
  +i(\omega_n-\omega_m)\langle\hat a_n^\dagger \hat a_m\rangle
  +L_{m,n}
  \end{eqnarray}
and
  \begin{eqnarray}
    \label{eq:acorrb}
    \frac{d}{dt}\langle\hat a_m \hat a_n^\dagger\rangle=
    \frac{1}{2}\sum_k\left(
    L_{k,m}-\Gamma_{k,m}\right)\langle a_k \hat a_n^\dagger\rangle
  +\frac{1}{2}\sum_k\left(
    L_{n,k}-\Gamma_{n,k}\right)\langle a_m \hat a_k^\dagger\rangle
  \nonumber\\
  +i(\omega_n-\omega_m)\langle\hat a_m \hat a_n^\dagger\rangle
  +\Gamma_{m,n}.
  \end{eqnarray}
\end{mathletters}
For the quasi-mode operators, we find the correlations to be
\begin{mathletters}
  \label{eq:Acorr}
  \begin{equation}
    \frac{d}{dt}\langle\hat A_\nu^\dagger \hat A_\mu\rangle=
    \left\{\frac{1}{2}(\lambda_\nu+\lambda_\mu-\gamma_\nu-\gamma_\mu)
      +i(\Omega_\nu-\Omega_\mu)\right\}
    \langle\hat A_\nu^\dagger \hat A_\mu\rangle
    +\frac{1}{\mathcal{E}_\nu\mathcal{E}_\mu}
    \sum_{n,m}L_{m,n}\varepsilon_n\varepsilon_m c^{(\nu)*}_n c^{(\mu)}_m
  \end{equation}
  and
  \begin{equation}
    \frac{d}{dt}\langle\hat A_\mu\hat A_\nu^\dagger\rangle=
    \left\{\frac{1}{2}(\lambda_\nu+\lambda_\mu-\gamma_\nu-\gamma_\mu)
      +i(\Omega_\nu-\Omega_\mu)\right\}
    \langle\hat A_\mu\hat A_\nu^\dagger \rangle
    +\frac{1}{\mathcal{E}_\nu\mathcal{E}_\mu}
    \sum_{n,m}\Gamma_{m,n}\varepsilon_n\varepsilon_m c^{(\nu)*}_n c^{(\mu)}_m.
  \end{equation}
\end{mathletters}


\begin{thebibliography}{99}
\bibitem{schawlow} A.\ L.\ Schawlow and C.\ H.\ Townes,
  Phys.\ Rev.\ {\bf 112}, 1940 (1958).
\bibitem{scully} See, e.g., M.\ Sargent III, M.\ O.\ Scully, and W.\ E.\ Lamb,
  Jr., {\em Laser Physics} (Addison-Wesly, Reading, MA 1974).
\bibitem{lax} See, e.g., M.\ Lax, in {\em Physics of Quantum
    Electronics}, ed.\ P.\ L.\ Kelley, B.\ Lax, and P.\ E.\ Tannenwald
  (McGraw-Hill, New York, 1966), p.\ 735.
\bibitem{peter} K.\ Petermann, IEEE J.\ Quantum Electron.\ {\bf
    QE-15}, 566 (1979).
\bibitem{output} W.\ A.\ Hamel and J.\ P.\ Woerdman,
  Phys.\ Rev.\ Lett.\ {\bf 64}, 1506 (1990).
\bibitem{semicond} Y.-J.\ Cheng, P.\ L.\ Musche, and A.\ E.\ Siegman,
  IEEE J.\ Quantum Electron.\ {\bf QE-30}, 1498 (1994);
  Y.-J.\ Cheng, G.\ Fanning, and A.\ E.\ Siegman,
  Phys.\ Rev.\ Lett.\ {\bf 77}, 627 (1996).
\bibitem{gas} M.\ A.\ van Eijkelenborg, \AA.\ M.\ Lindberg, M.\ S.\
  Thijssen, and J.\ P.\ Woerdman, Phys.\ Rev.\ Lett.\ {\bf 77}, 4314
  (1996);
  M.\ A.\ van Eijkelenborg, \AA.\ M.\ Lindberg, M.\ S.\
  Thijssen, and J.\ P.\ Woerdman, Opt.\ Commun.\ {\bf 137}, 303
  (1997);
  M.\ A.\ van Eijkelenborg, M.\ P.\ van Exter, and J.\ P.\ Woerdman,
  Phys.\ Rev.\ A {\bf 57}, 571 (1998).
\bibitem{polar} A.\ M.\ van der Lee, N.\ J.\ van Druten, A.\ Mieremet,
  M.\ A.\ van Eijkelenborg, \AA.\ M.\ Lindberg, M.\ P.\ van Exter, and
  J.\ P.\ Woerdman, Phys.\ Rev.\ Lett.\ {\bf 79}, 4357 (1997);
  O.\ Emile, M.\ Brunel, A.\ Le Floch, and F.\ Bretenaker,
  Europhys.\ Lett.\ {\bf 43}, 153 (1998).
\bibitem{aperture} \AA.\ M.\ Lindberg, M.\ A.\ van Eijkelenborg, K.\
  Joosten, G.\ Nienhuis, and J.\ P.\ Woerdman,
  Phys.\ Rev.\ A {\bf 57}, 3036 (1998);
  O.\ Emile, M.\ Brunel, F.\ Bretenaker, and A.\ Le Floch,
  Phys.\ Rev.\ A {\bf 57}, 4889 (1998).
\bibitem{color} A.\ M.\ van der Lee, M.\ P.\ van Exter, A.\ L.\
  Mieremet, N.\ J.\ van Druten, and J.\ P.\ Woerdman,
  Phys.\ Rev.\ Lett.\ {\bf 81}, 5121 (1998).
\bibitem{semi} A.\ E.\ Siegman, Phys.\ Rev.\ A {\bf 39}, 1253 (1989),
  and references therein; A.\ E.\ Siegman, submitted.
\bibitem{squant} Ph.\ Goldberg, P.\ W.\ Milonni, and B. Sundaram,
  Phys.\ Rev.\ A {\bf 44}, 1969 (1991);
  Ph.\ Grangier and J.-Ph.\ Poizat, Eur.\ Phys.\ D {\bf
    1}, 97 (1998);
  C.\ Lamprecht and H.\ Ritsch, submitted.
\bibitem{bardroff} P.\ J.\ Bardroff and S.\ Stenholm, submitted to
  Phys.\ Rev.\ Lett.
\bibitem{lang} R.\ Lang, M.\ O.\ Scully, and W.\ E.\ Lamb,
  Phys.\ Rev.\ A {\bf 7}, 1788 (1973).
\bibitem{fn:ev} For the case of a symmetric matrix $L_{m,n}-\Gamma_{m,n}$, the
  relation between left and right eigenvectors is easy to see.
  However, a similar relation holds also in the case of complex mode
  functions $u_n(x)$, when the matrix $L_{m,n}$ is Hermitian.
  The left eigenvectors are then given by
  $\sum_m M_{n,m} \varepsilon^2_m c^{(\nu)}_m$ with a unitary matrix
  $M_{n,m}$.
\bibitem{footnote} This is true as long as the non-Hermitian matrix has no
  degenerate eigenvalues. For a discussion of matrix properties see,
  e.g.,
  P.\ Lencaster, {\em Theory of Matrices}, (Academic Press, New York,
  1969).
\end{thebibliography}
\end{document}